# Room temperature magnetization switching in topological insulator-ferromagnet heterostructures by spin-orbit torques

Yi Wang[1*], Dapeng Zhu[1*], Yang Wu[1], Yumeng Yang[1], Jiawei Yu[1], Rajagopalan Ramaswamy[1], Rahul Mishra[1], Shuyuan Shi[1,2], Mehrdad Elyasi[1], Kie-Leong Teo[1], Yihong Wu[1], and Hyunsoo Yang[1,2]

[1]Department of Electrical and Computer Engineering, National University of Singapore, 117576, Singapore
[2]Centre for Advanced 2D Materials, National University of Singapore, 6 Science Drive 2, 117546 Singapore

[*]These authors contributed equally to this work.
Correspondence and requests for materials should be addressed to H.Y. (email: eleyang@nus.edu.sg).

Topological insulators (TIs) with spin momentum locked topological surface states (TSS) are expected to exhibit a giant spin-orbit torque (SOT) in the TI/ferromagnet systems. To date, the TI SOT driven magnetization switching is solely reported in a Cr doped TI at 1.9 K. Here, we directly show giant SOT driven magnetization switching in a $Bi_2Se_3$/NiFe heterostructure at room temperature captured using a magneto-optic Kerr effect microscope. We identify a large charge-to-spin conversion efficiency of ~1–1.75 in the thin TI films, where the TSS is dominant. In addition, we find the current density required for the magnetization switching is extremely low, ~$6\times10^5$ A $cm^{-2}$, which is one to two orders of magnitude smaller than that with heavy metals. Our demonstration of room temperature magnetization switching of a conventional 3$d$ ferromagnet using $Bi_2Se_3$ may lead to potential innovations in TI based spintronic applications.

The spin currents generated by charge currents via the spin Hall effect[1-4] and/or Rashba-Edelstein effect[5,6] can exert spin-orbit torques (SOTs) on the adjacent FM layer and result in the current induced magnetization switching. A higher charge-to-spin conversion efficiency (referred as SOT efficiency) is crucial for the low power dissipation SOT applications. Recently, the SOTs have been studied in topological insulators (TIs)[7-14], which are an emerging state of quantum matter possessing spin-momentum-locked topological surface states (TSS)[15-17]. This exotic property is supposed to exhibit a large SOT efficiency, which is explored recently by the spin transport methods, such as spin torque ferromagnetic resonance (ST-FMR)[7,8,13], spin pumping[9,10,14,18], and spin tunneling spectroscopy[19,20]. However, in TIs such as $Bi_2Se_3$, the bulk states (BS) and two dimensional electron gas (2DEG), which are typically present due to defects in the bulk and band bending at the surface[21,22], can lead to an inevitable contamination to the SOT effects from TSS. This is indicated by a wide range of the SOT efficiencies of 0.01–3.5 reported in the $Bi_2Se_3$/ferromagnet (FM) systems[7-9,18]. The roles of BS, 2DEG, and TSS on SOT efficiencies have not yet been clearly understood in details, which is critical for highly efficient SOT driven magnetization switching using TIs.

To date, the magnetization switching induced by TI SOT is solely reported in a Cr doped TI at a very low temperature (1.9 K) with an external magnetic field[11], and the SOT induced magnetization switching in a TI/3$d$ FM heterostructure at room temperature is highly desired for applications. Here, we obtain a TSS dominated SOT effect in 5–8 quintuple layers (QL) of $Bi_2Se_3$ films, exhibiting a large SOT efficiency of ~1–1.75 at room temperature using ST-FMR measurements. By taking advantage of the high efficiency, we image the SOT induced magnetization switching by a magneto-optic Kerr effect (MOKE) microscope in the $Bi_2Se_3$/NiFe (Py) heterostructures at room temperature for the first time after injecting a pulsed dc current.

The required current density for SOT switching is extremely low and is one to two orders of magnitude smaller than that with heavy metals[23-25]. Our results suggest that TI/FM heterostructure could be a potential candidate for room temperature spintronic devices with ultralow power dissipation.

High quality $Bi_2Se_3$ films ranging from 5 to 20 QL (1 QL ≈ 1 nm) were grown on $Al_2O_3$ (0001) substrates using molecular beam epitaxy (MBE) technique (see Methods). Figure 1a shows the atomic-force microscopy (AFM) image of a representative 10-QL $Bi_2Se_3$ film, indicating a smooth surface and high film quality. From the four-probe and Hall measurements at room temperature, we find that the resistivity ($\rho_{BiSe}$) is ~1,000 μΩ cm at large thicknesses (15 and 20 QL), increases at 10 QL and becomes ~4,117 μΩ cm at 5 QL, as shown in Fig. 1b. Moreover, the sheet resistance shows a similar trend as $\rho_{BiSe}$ (see Supplementary Note 2 and Supplementary Fig. 2). The sheet carrier concentration ($n_{2D}$) shows an opposite trend, decreasing from ~$6\times10^{13}$ cm$^{-2}$ at 20 QL to ~$3.8\times10^{13}$ cm$^{-2}$ below 8 QL. This behavior suggests a small contribution of BS and 2DEG to electrical transport properties in the thin $Bi_2Se_3$ cases as we discuss later. We also characterize $Bi_2Se_3$ thickness ($t_{BiSe}$) dependent $\rho_{BiSe}$ (Fig. 1c) and $n_{2D}$ (Fig.1d) at different temperatures. Our $Bi_2Se_3$ films show a typical metallic behavior similar to previous reports[8,26].

Figure 2a shows the schematic diagram of the ST-FMR measurement (see Methods), an effective technique to evaluate the SOT efficiency[7,27]. The ST-FMR devices consist of $Bi_2Se_3$ ($t_{BiSe}$)/$Co_{40}Fe_{40}B_{20}$ (CFB, 7 nm) bilayers. Figure 2b illustrates the current induced spin polarization and magnetization dynamics in $Bi_2Se_3$/CFB bilayers. As an in-plane rf current ($I_{RF}$) flows in the $Bi_2Se_3$ layer, non-equilibrium spins are generated at the $Bi_2Se_3$ surfaces denoted by the arrows with green and red balls. These spins from $Bi_2Se_3$ top surface diffuse into CFB and exert oscillating damping-like torque ($\tau_{DL}$) and/or a field-like torque ($\tau_{FL}$) on the magnetization. These torques together with rf current induced Oersted field ($H_{RF}$) torque ($\tau_{Oe}$) trigger the precession of CFB magnetization and an oscillation of the anisotropic magnetoresistance with the same frequency as $I_{RF}$. Consequently, a mixing dc voltage $V_{mix}$ (i.e. ST-FMR signal) is produced across the ST-FMR device[7,8,27,28].

Figure 2c shows typical ST-FMR signals $V_{mix}$ (open symbols), which are fitted by $V_{mix} = V_S F_S + V_A F_A$, where $F_S$ and $F_A$ are symmetric and antisymmetric Lorentzian functions, respectively. The amplitude of symmetric ($V_S$) and antisymmetric component ($V_A$) are attributed to $\tau_{DL}$ and $\tau_{FL} + \tau_{Oe}$, respectively[7,27]. By adopting the established analysis method[7,8], the SOT efficiency ($\theta_{TI} = J_S/J_C$) can be evaluated from only $V_S$ (see Supplementary Note 3), where $J_S$ is the spin current density at the Bi$_2$Se$_3$/CFB interface and $J_C$ (A cm$^{-2}$) is the uniform charge current density in the Bi$_2$Se$_3$ layer. Similar measurements and analyses are further performed on devices with various $t_{BiSe}$ spanning 5–20 QL. Figure 3 shows $\theta_{TI}$ versus $t_{BiSe}$ at room temperature. Each data point is averaged from three devices, which show a similar behavior. Specifically, $\theta_{TI}$ shows a constant value of ~0.3 for 15 and 20 QL devices, and starts to increase below 10 QL, reaching a maximum of ~1.75 at 5 QL. $\theta_{TI}$ in thinner films increases more than 5 times compared to that in thicker Bi$_2$Se$_3$ devices. From the line shape of the ST-FMR signals and the positive sign of $\theta_{TI}$ at different $t_{BiSe}$, we confirm that the direction of in-plane spin polarization ($S_\parallel$) at the interface of Bi$_2$Se$_3$ and CFB is in line with TSS where $S_\parallel$ is locked at right angles to the electron momentum[7,8,29-31].

Recent experimental and theoretical works[21,32] indicate that as the thickness of Bi$_2$Se$_3$ decreases to several QLs, BS shrink significantly and finally disappear. In addition, the surface 2DEG bands are gradually quantized into discrete subbands enclosed by the linear TSS bands due to quantum confinement effects (see Supplementary Fig. 3). Since the thickness of a TSS ($t_{TSS}$) and 2DEG ($t_{2DEG}$) in $Bi_2Se_3$ are reported to be ~1 nm[21,26,32,33] and ~4 nm[21,26,32], respectively, negligible BS are expected when the $Bi_2Se_3$ thickness is less than 8 QL. Accordingly, we discuss the transports in three regions (I, II and III, denoted by different colours) in Fig. 3. In region I ($t_{BiSe}$ > 10 QL), there are considerable BS and 2DEG contributions to the transport, which could dilute the TSS[8], resulting in a small $\theta_{TI}$. In region II (~10 QL), BS start to shrink, leading to a slight incrase of $\theta_{TI}$. In region III ($t_{BiSe} \leq 8$ QL), the BS disappear and the contribution from the 2DEG decreases as we discuss later. On the other hand, due to the lack of inversion symmetry in our devices, Rashba splitting states in 2DEG subbands can give rise to $S_\parallel$. However, the accumulated spins due to the Rashba states are expected to have an opposite helicity (i.e. negative $\theta_{TI}$) compared to the TSS[21,34-36]. Since $\theta_{TI}$ always shows positive values in all our devices, we conclude that the TSS dominated SOT is the main contribution to the large enhancement of $\theta_{TI}$ in region III.

To further confirm that the TSS dominate SOT in region III (5–8 QL), we establish a model to quantify the carrier concentration in TSS ($n_{TSS}$), 2DEG ($n_{2DEG}$) and BS ($n_{2D-Bulk}$), as well as the corresponding current shunting effect due to BS and 2DEG (see Supplementary Note 4 and Supplementary Fig. 3). As shown in Fig. 4a, the $n_{2DEG}$ decreases significantly as $t_{BiSe} \leq 8$ QL, while $n_{TSS}$ shows a slight increase as $t_{BiSe}$ decreases. This observation reproduces the inherent behaviors of TSS and 2DEG carriers measured in very thin $Bi_2Se_3$ films[26]. Moreover, the larger value of $n_{TSS}$ compared to $n_{2DEG}$ for $t_{BiSe} \leq 8$ QL corroborates a TSS dominated transport in thin $Bi_2Se_3$ film region (see Supplementary Note 4 and Supplementary Fig. 4-7). Figure 4b shows the location of the Fermi level ($E_F$) relative to Dirac point ($E_{DP}$) and Fermi vector ($k_F$), we find that $E_F - E_{DP}$ ($k_F$) gradually increases from ~403 to 447 meV (from ~0.123 to 0.135 Å$^{-1}$) as $t_{BiSe}$ decreases, indicating that the DP slightly moves downwards to a larger binding energy which accounts for the weak increase of $n_{TSS}$. The value of $E_F - E_{DP}$ ($k_F$) and the DP movement are in line with previous ARPES measurements for $Bi_2Se_3$ films[35,37-40]. Figure 4c shows that the charge currents in the TSS on the top surface over the total currents flowing in $Bi_2Se_3$ ($I_{TSS}/I_{total}$) increases from ~0.2 to 0.4 as $t_{BiSe}$ decreases from 20 to 5 QL (see Supplementary Note 5), which again verifies that the TSS dominates the region III.

In addition, we estimate the *interface* SOT efficiency from TSS ($\lambda_{TSS}$) by using an interface charge current density $J_{C\text{-}TSS}$ (A cm$^{-1}$) in TSS (see Supplementary Note 6 and Supplementary Fig. 8-11). As shown in Fig. 4d, we find $\lambda_{TSS}$ is in the range of ~0.38–0.82 nm$^{-1}$ when $t_{BiSe} \leq 8$ QL at room temperature, which is consistent with recently reported interface SOT efficiency values in $(Bi_{1-x}Sb_x)_2Te_3$[13]. In principle, $\lambda_{TSS}$ is inversely propotional to the Fermi velocity $V_F$ and remains almost constant at different $t_{BiSe}$,[13] however a pronounced variation of $\lambda_{TSS}$ is observed. This deviation unambiguously suggests that there is an opposite spin accumulation mechanism which cancels part of the spins generated by TSS in $Bi_2Se_3$. We attribute this to the Rashba states in 2DEG[21,34-36]. From the change of $\lambda_{TSS}$, we can extract the interface SOT efficiency from 2DEG ($\lambda_{2DEG}$) $\approx -0.4$ nm$^{-1}$ in the thin film regime (see Supplementary Note 7). After excluding the 2DEG contribution, the amended interface SOT efficiency from TSS denoted by red circles in Fig. 4d, shows a constant value of ~0.8 nm$^{-1}$ for 7, 8 and 10 QL $Bi_2Se_3$ devices, which is simliar to the value of $\lambda_{TSS}$ ~0.82 nm$^{-1}$ at $t_{BiSe} = 5$ QL (see Supplementary Note 7 and Supplementary Fig. 8-11).

The ST-FMR measurements and the above analysis reveal that the contribution of TSS is dominant in the thin $Bi_2Se_3$ films (5–8 QL), leading to a higher SOT efficiency at room temperature. Subsequently, we demonstrate the SOT induced magnetization switching in $Bi_2Se_3$ (8 QL)/Py (6 nm) heterostructures (see Methods) at room temperature by applying a pulsed dc current $I$. The high-resolution scanning transmission electron microscope (STEM) image shows a clean and smooth interface between the $Bi_2Se_3$ and Py layer (see Supplementary Note 1 and Supplementary Fig. 1). In order to take advantage of the higher SOT efficiency and flow enough charge currents in the $Bi_2Se_3$ layer, an 8-QL $Bi_2Se_3$ is utilized. As depicted in Fig. 5a, the continuous Py layer is separated into five well-defined rectangles (yellow dashed boxes) and magnetically isolated by Cu bars. The magnetic easy axis of Py rectangles is along $\pm y$ directions due to the shape anisotropy. The magnetization direction of Py is collinear with the incoming spin directions (see Fig. 2b) and thus the spins can directly switch the magnetization direction of Py without any external assisted magnetic field, which are captured by MOKE imaging measurements (see Methods).

Figure 5a-e (top panel) show the SOT driven magnetization switching by applying a pulsed $I$ along the +$x$-axis. At the beginning of this set of measurements, we first saturate the Py magnetization along the +$y$-axis by applying an in-plane external magnetic field ($H$). Then we remove $H$ and apply $I$ along the +$x$-axis to the device. When the current density in $Bi_2Se_3$ ($J_C$) is zero, we capture the MOKE image as shown in Fig. 5a. The dark contrast represents the magnetization along the +$y$-axis, indicated by the white arrow. We find that as $J_C$ increases, the area of the switched magnetization with light contrast gradually expands (see Fig. 5b-d). Finally, the magnetization of all Py rectangles is switched to the −$y$-axis at $J_C = 5.7 \times 10^5$ A cm$^{-2}$, which is indicated by the white arrow in Fig. 5e. Similarly, for the other set of measurements in Fig. 5f-j, we first initialize the Py magnetization along the −$y$-axis. Then we remove $H$ and apply $I$ of opposite polarity, i.e., along the −$x$-axis. As $J_C$ increases, the Py magnetization switches from the −$y$ (Fig. 5f, light contrast) to +$y$-axis (Fig. 5j, dark contrast) at $J_C = 6.2 \times 10^5$ A cm$^{-2}$, exhibiting

the opposite switching direction. The SOT induced switching is reproducible in other devices (see Supplementary Note 8 and Supplementary Fig. 12). We find that the current density required for the room temperature SOT induced magnetization switching in $Bi_2Se_3$/Py is extremely low at ~$6 \times 10^5$ A cm$^{-2}$, which is one to two orders of magnitude smaller than that with heavy metals[23-25]. Moreover, based on the conventional antidamping spin torque driven magnetization switching model[24,41] with consideration of thermal fluctuation and reverse domain nucleation, we determine the SOT efficiency for $Bi_2Se_3$/Py to be ~1.71, which is in accord with the value from our ST-FMR measurements (see Supplementary Note 9). This agreement further corroborates the excellent efficiency of TIs in spin generation and SOT driven magnetization switching. Moreover, the robust SOT induced magnetization switching is also observed in devices with a Cu or NiO insertion layer between the $Bi_2Se_3$ and Py layer (see Supplementary Note 11 and Supplementary Fig. 14-16). From the control measurements, we find that neither the Joule heating nor the current induced Oersted field could lead to the observed current induced magnetization switching (see Supplementary Fig. 13 and Supplementary Fig. 17).

The fundamental obstacle for high-density non-volatile applications of magnetic devices in a conventional spin torque scheme is the high critical switching current density, resulting in a large size of the current driving transistor. Utilizing the giant SOT effect in $Bi_2Se_3$, which can be grown in a wafer scale using MBE, we achieve a significantly low $J_C$ to switch a conventional 3$d$ ferromagnet NiFe, which is widely utilized in industries, addressing an outstanding scalability issue in modern magnetic devices. In addition, no requirement of an assistive magnetic field for our demonstrated magnetization switching scheme makes the TI/FM material systems easy to integrate into the established industrial technology for magnetic devices. Our above findings may bring this exotic newly discovered quantum matter from research activities to core ingredients in real spintronic applications.

**Methods**

**Film growth and device fabrication.** High quality $Bi_2Se_3$ films ranging from 5 to 20 QLs are grown on $Al_2O_3$ (0001) substrates in a MBE system (MBC-1000-2C from ULVAC) with a base pressure $< 1.5 \times 10^{-9}$ Torr, by using two-step deposition procedure[42,43]. The sapphire substrates are first cleaned in acetone, isopropanol, and de-ionized water, and subsequently annealed at 750 °C for 30 min in a vacuum after being transferred into the growth chamber. Elemental Bi (6N) and Se (5N) solid sources are evaporated from standard Knudsen cells under a Se/Bi flux ratio of ~20. To reduce Se vacancies in $Bi_2Se_3$, initial 2–3 QL $Bi_2Se_3$ are deposited at 150 °C, and then the substrate temperature is ramped to 250 °C at 5 °C/min under Se flux for the second step growth. Our $Bi_2Se_3$ films have smooth surface with a roughness of ~0.5 nm and show a clear terrace step of ~1 nm ($\approx$ 1 QL). The morphology indicates the high quality of our $Bi_2Se_3$ films. After the $Bi_2Se_3$ growth, the bare $Bi_2Se_3$ films are immediately transferred into a magnetron sputtering chamber via air in the standard cleanroom environment with a well-controlled levels of low humidity and constant temperature. The transfer time was strictly controlled under 5 minutes before pumping down the sputtering chamber. For the ST-FMR devices, a 7-nm thick $Co_{40}Fe_{40}B_{20}$ (CFB) is subsequently sputtered on the $Bi_2Se_3$ film with a low power of 60 W at room temperature with a base pressure of ~$3 \times 10^{-9}$ Torr. Finally, the $Bi_2Se_3$/CFB bilayer is protected by the sputtered MgO (2 nm)/$Al_2O_3$ (3 nm) layer. For the MOKE imaging devices, the Py (6 nm)/MgO (1 nm)/$SiO_2$ (4 nm) stacks are subsequently sputtered onto the $Bi_2Se_3$ (8 QL) films with an *in-situ* magnetic field along the *y*-axis (i.e. perpendicular to the current channel, see Fig. 5f) at room temperature with a base pressure of ~$3 \times 10^{-9}$ Torr. A very low sputtering power of 40 W is used for the Py deposition. Subsequently, five 2-µm wide grooves on the Py layer are etched and backfilled with nonmagnetic metal Cu, which divide the continuous Py layer into five rectangles and make them magnetically isolated. All devices are patterned by photolithography and ion milling.

**ST-FMR measurements.** The ST-FMR signals are detected by a lock-in amplifier. The frequencies and nominal power of the rf current $I_{RF}$ are 6–9 GHz and 15 dBm, respectively. An in-plane external magnetic field ($H$) is applied at a fixed angle ($\theta_H$) of 35 ° with respect to $I_{RF}$.

**MOKE imaging measurements.** The Py magnetic easy anisotropy in the device is along the ±$y$-axis (see Fig. 5f) due to the shape anisotropy. This allows us to capture the magnetization switching after pulsed dc current is off, where there is no current induced spurious effects in the MOKE images. For the MOKE imaging measurements, we first saturate the Py magnetization along the +$y$ or −$y$-axis with an in-plane external magnetic field $H$, then we remove $H$ and apply a pulsed dc current (500 µs pulse width) to observe the magnetization switching using MOKE microscope.

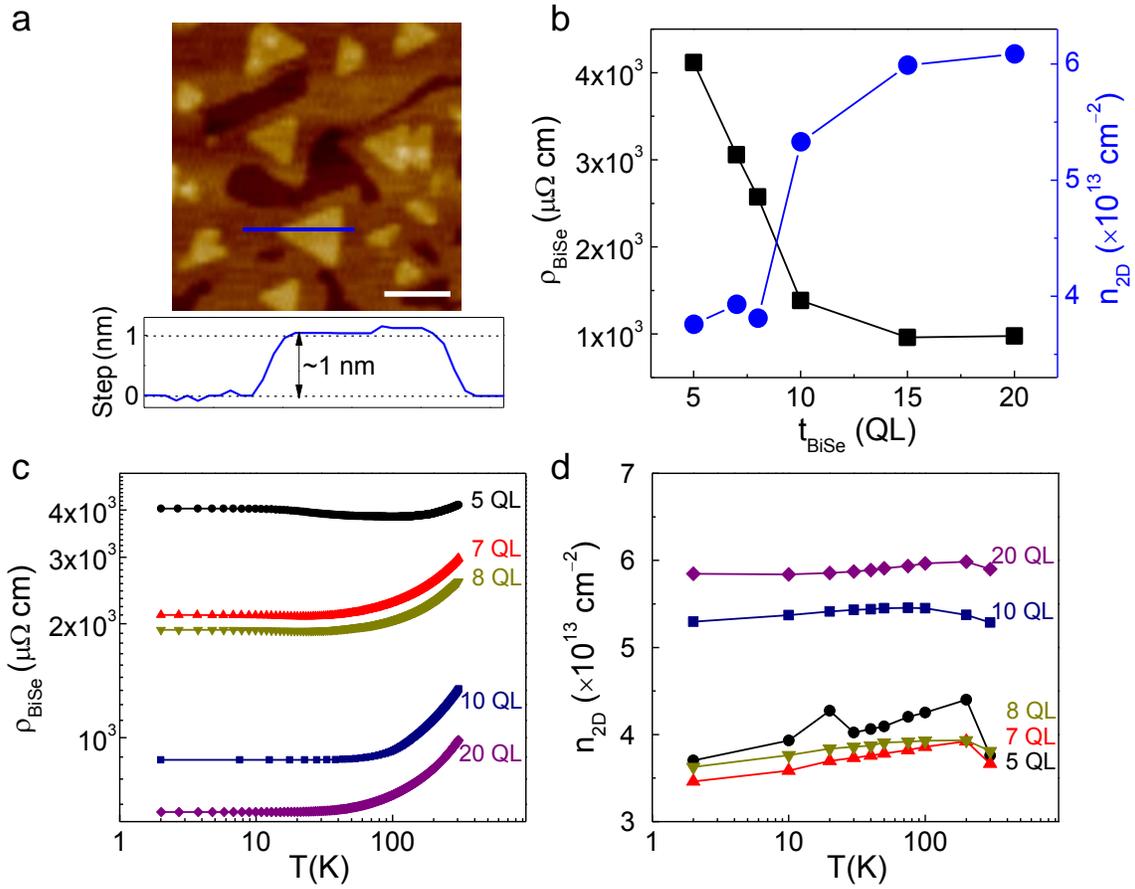

**Figure 1 | Bi$_2$Se$_3$ film properties. a**, AFM image of a 10-QL Bi$_2$Se$_3$ film with a roughness of ~0.5 nm. 1 QL (≈ 1 nm) step of crystal terrace along the blue line is clearly seen in the bottom graph. The white scale bar is 100 nm. **b**, Bi$_2$Se$_3$ thickness dependent resistivity, $\rho_{BiSe}$, and sheet carrier concentration, $n_{2D}$, in Bi$_2$Se$_3$ at room temperature. **c-d**, $\rho_{BiSe}$ and $n_{2D}$ as a function of temperature for Bi$_2$Se$_3$ films with different thicknesses, respectively.

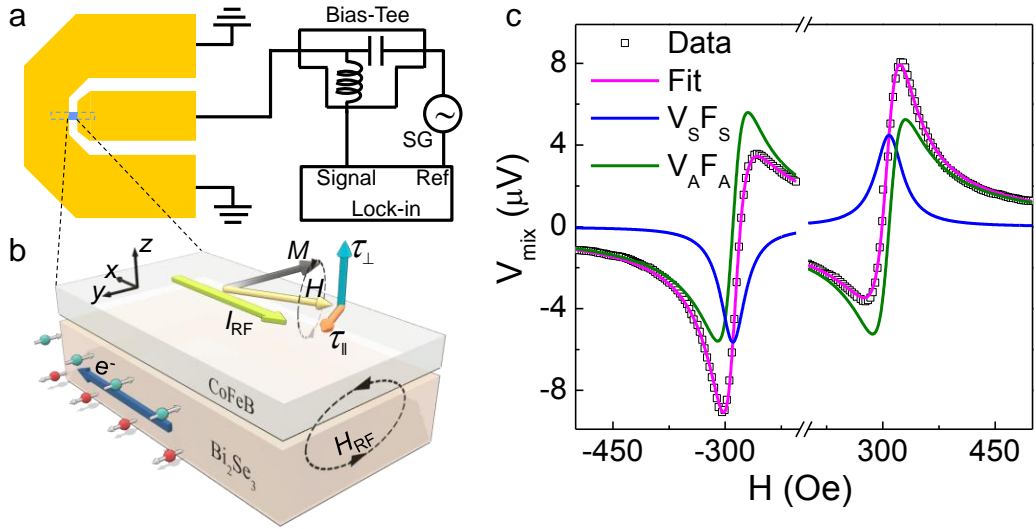

**Figure 2 | ST-FMR measurements and signals with fits. a**, Schematic diagram of the ST-FMR measurement set-up, illustrating a ST-FMR device and the measurement circuit. An rf current from a signal generator (SG) is injected into the ST-FMR devices via a bias-tee. **b**, Illustration of spin momentum locking and SOT induced magnetization dynamics in the ST-FMR measurements. The big blue arrow denotes the electron moving direction (opposite to $I_{RF}$ direction). The arrows with green and red balls denote the spin polarizations generated at top and bottom surfaces of $Bi_2Se_3$, respectively. **c**, A typical ST-FMR signal (open symbols) from a $Bi_2Se_3$ 20 QL/CFB 7 nm device at 6 GHz with fits (solid lines), where the blue and green lines represent the symmetric Lorentzian ($V_S F_S$) and antisymmetric Lorentzian ($V_A F_A$) components, respectively.

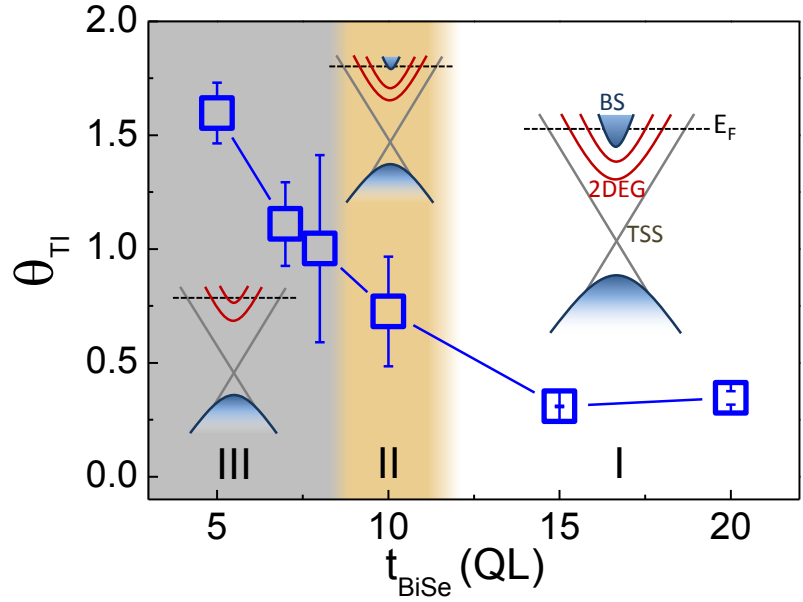

**Figure 3 | The SOT efficiency ($\theta_{TI}$) as a function of Bi$_2$Se$_3$ thickness ($t_{BiSe}$) at room temperature.** Each $\theta_{TI}$ represents the averaged value from three devices. The error bars are the standard deviation. Region I, II and III denoted by different colours represent the charge-to-spin conversion dominated by different mechanisms. The inset shows the schematic of the band structure for each region.

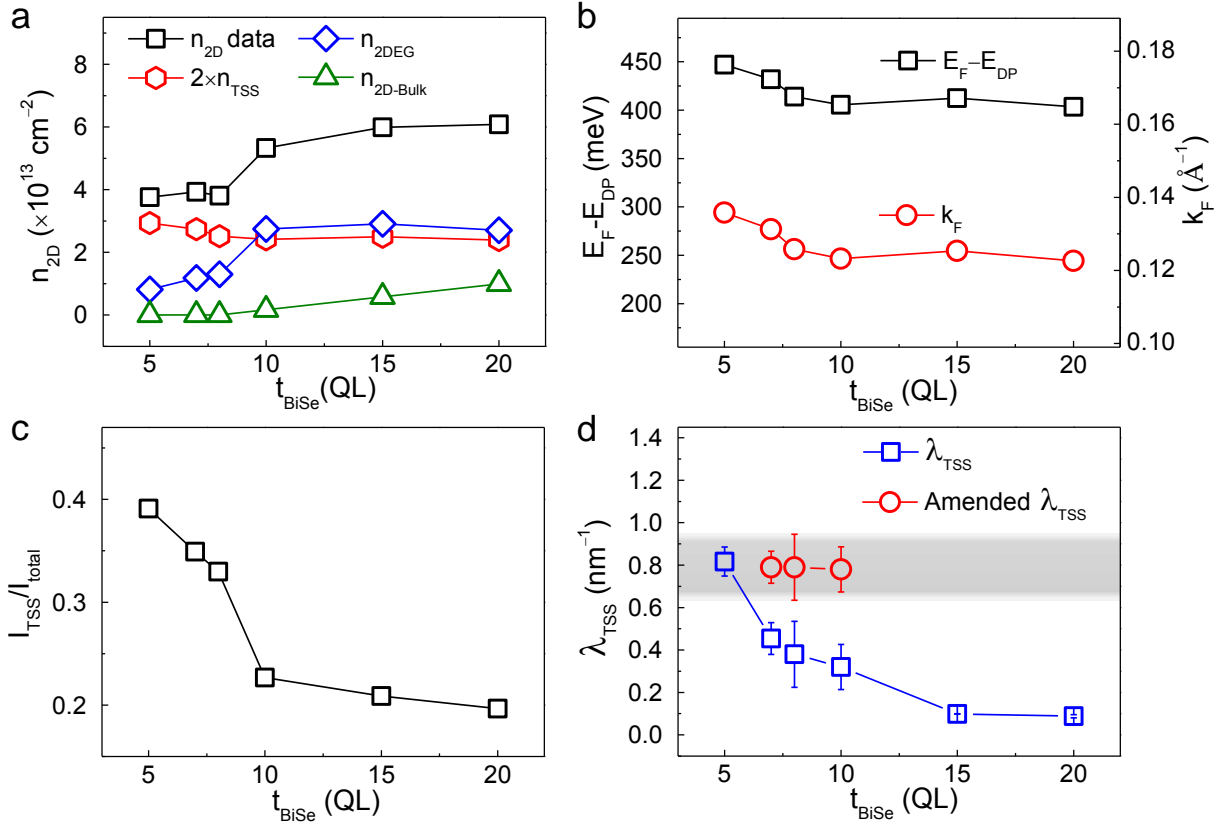

**Figure 4 | Model analysis results and TSS dominated SOT in 5-8 QL Bi$_2$Se$_3$. a**, Sheet carrier concentration of TSS, 2DEG and bulk channels. **b**, $E_F - E_{DP}$ and $k_F$ of TSS. **c**, Ratio of charge currents in the TSS on the top surface and total currents in Bi$_2$Se$_3$. **d**, Interface SOT efficiency, $\lambda_{TSS}$ (blue squares), as a function of $t_{BiSe}$ at room temperature. The amended interface SOT efficiency from TSS after excluding the opposite 2DEG contribution is shown for 7, 8 and 10 QL Bi$_2$Se$_3$ in (**d**) with red circles. The factor of 2 in (**a**) arises due to the consideration of both the bottom and the top TSS in Bi$_2$Se$_3$. The error bars in (**d**) are the standard deviation from three devices at each Bi$_2$Se$_3$ thickness.

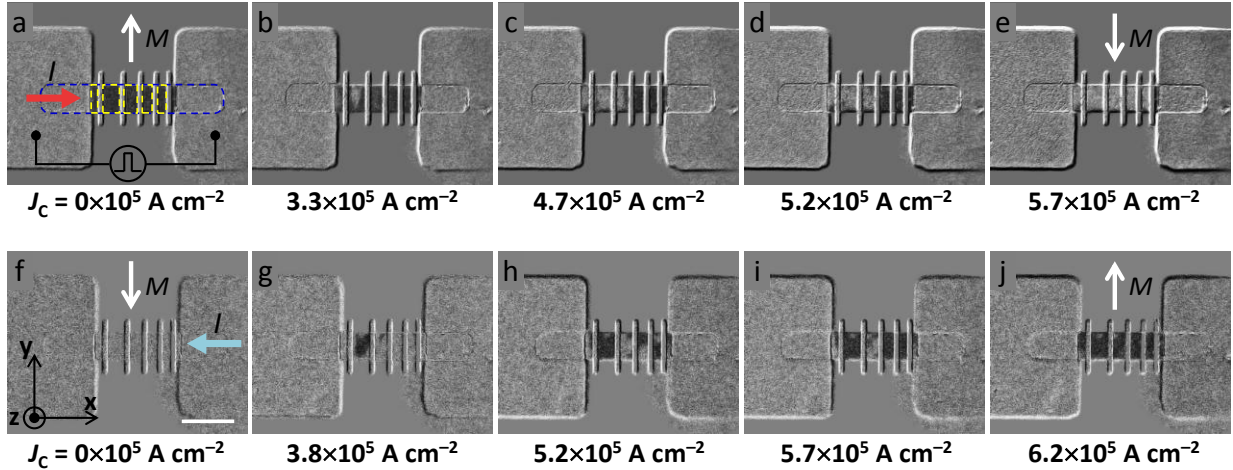

**Figure 5 | MOKE images of SOT driven magnetization switching in $Bi_2Se_3$/Py at zero magnetic field and room temperature. a-e**, MOKE images for SOT driven magnetization switching by applying a pulsed dc current $I$ along the +$x$-axis with increasing the current density $J_C$ in the $Bi_2Se_3$ layer denoted underneath the corresponding image. The blue dashed rectangle in (**a**) represents the 12-μm wide $Bi_2Se_3$/Py channel connected with two big contact pads. The yellow dashed boxes in (**a**) denote five small Py rectangles magnetically isolated by Cu bars. **f-j**, MOKE images for SOT driven magnetization switching for $I$ along –$x$-axis. The dark (light) contrast shows the magnetization along +$y$ (–$y$)-axis. The direction of magnetization is also indicated by the white arrows in (**a**), (**e**), (**f**) and (**j**). The white scale bar is 20 μm.